\documentclass{aa}
\usepackage{epsf}

\begin{document}
\title{Three types of cooling superfluid neutron stars: \\
        Theory and observations}
\author{
A. D. Kaminker \inst{1}
\and
D. G. Yakovlev\inst{1}
\and
O. Y. Gnedin\inst{2}
}
\institute{
Ioffe Physical Technical Institute,
         Politekhnicheskaya 26, 194021 St.~Petersburg, Russia 
\and
Space Telescope Science Institute,
         3700 San Martin Drive, Baltimore, MD 21218, USA
\\
{\em  kam@astro.ioffe.rssi.ru,  yak@astro.ioffe.rssi.ru,
ognedin@stsci.edu
}}
\offprints{A.D.\ Kaminker}

\date{Received x xxx 2001 / Accepted x xxx 2001}
\abstract{
Cooling of neutron stars (NSs) with the cores composed of neutrons, 
protons, and electrons is simulated assuming $^1$S$_0$ pairing of 
neutrons in the NS crust, and also $^1$S$_0$ pairing of protons 
and weak $^3$P$_2$ pairing of neutrons in the NS core, and using 
realistic density profiles of the superfluid critical temperatures
$T_{\rm c}(\rho)$. The theoretical cooling models of isolated
middle-aged NSs can be divided into three main types.  
(I) {\it Low-mass}, {\it slowly cooling} NSs where the direct 
Urca process of neutrino emission is either forbidden  
or almost fully suppressed by the proton superfluidity.
(II) {\it Medium-mass} NSs which show {\it moderate}
cooling via the direct Urca process suppressed by
the proton superfluidity. (III) {\it Massive} NSs which show 
{\it fast} cooling via the direct Urca process weakly suppressed by
superfluidity. Confronting the theory with observations 
we treat RX J0822--43, PSR 1055--52 and RX J1856--3754
as slowly cooling NSs. To explain these sufficiently warm sources
we need a density profile $T_{\rm c}(\rho)$ 
in the crust with a rather high and flat maximum and sharp wings.  
We treat 1E 1207--52, RX J0002+62, PSR 0656+14, Vela, and
Geminga as moderately cooling NSs. We can determine
their masses for a given model of proton superfluidity, 
$T_{\rm cp}(\rho)$, and the equation of state in the NS core. 
No rapidly cooling NS has been observed so far. 
}
\titlerunning{Three types of   
cooling superfluid neutron stars}
\authorrunning{A.~D.\ Kaminker, D.~G.\ Yakovlev, \& O.~Y.\ Gnedin}
\maketitle

\section{Introduction}
\label{sect-intro}
Cooling of neutron stars (NSs) depends on the properties of matter
of subnuclear and supranuclear density in the NS crusts and
cores.  These properties are still poorly known
and cannot be predicted precisely by contemporary
microscopic theories. 
For instance, microscopic calculations
of the equation of state (EOS) of matter in the NS cores
(e.g., Lattimer \& Prakash \cite{lp01})
or the superfluid properties of NS cores and crusts
(e.g., Lombardo \& Schulze \cite{ls01}) show a large
scatter of results depending on a model of strong interaction
and a many-body theory employed. 
It is important that these properties can be studied by confronting
the results of simulations of NS cooling with the observations
of thermal emission from isolated middle-aged NSs.

This paper is devoted to such studies. For simplicity,
we use the NS models with the cores composed of neutrons (n)
with an admixture of protons (p) and electrons. We will mainly focus on
the superfluid properties of NS matter which are characterized by
the density-dependent critical temperatures $T_{\rm c}(\rho)$ 
of nucleons. It is customary to consider superfluidities of three types:
singlet-state ($^1$S$_0$) superfluidity ($T_{\rm c}=T_{\rm cns}$)
of neutrons in the inner
NS crust and the outermost core; $^1$S$_0$ proton
superfluidity in the core ($T_{\rm c}=T_{\rm cp}$);
and triplet-state ($^3$P$_2$) neutron superfluidity in 
the core ($T_{\rm c}=T_{\rm cnt}$). Superfluidity of nucleons suppresses 
neutrino processes involving nucleons and affects
nucleon heat capacities (e.g., Yakovlev et al.\ \cite{yls99}).
In addition, it initiates a specific mechanism of neutrino
emission associated with Cooper pairing of nucleons
(Flowers et al.\ \cite{frs76}). 
Our aim is to analyze which critical temperatures
$T_{\rm c}(\rho)$ are consistent with observations
and do not contradict the current microscopic calculations.

We have considered this problem in two prior publications.
Kaminker et al.\ (\cite{khy01}, hereafter Paper I) analyzed
the effects of proton superfluidity (basing on one
particular model of $T_{\rm cp}(\rho)$) and $^3$P$_2$ 
neutron superfluidity in the NS core. 
Yakovlev et al.\ (\cite{ykg01}, hereafter
Paper II) included, additionally, 
the effects of $^1$S$_0$ neutron superfluidity
in the crust. Calculations in Papers I and II were performed 
for one particular EOS in the NS core. 
In the present paper we extend the results
of Papers I and II by considering three models of proton superfluidity
and another EOS in the NS core. 
We combine the results of Papers I and II
and give an overall analysis of the problem. We show
that one can distinguish
{\it three} distinctly different types of isolated 
middle-aged NSs, which show {\it slow}, {\it moderate}, and {\it fast}
cooling.
Using this concept
we discuss a possible interpretation of observations
of thermal emission from eight middle-aged isolated NSs.
 
\section{Cooling models}
\label{sect-model}

We simulate NS cooling with our fully relativistic
non-isothermal cooling code (Gnedin et al.\ \cite{gyp01}). 
The code solves the radial heat diffusion equation 
in the NS interior (excluding the outer heat-blanketing
layer placed at $\rho < 10^{10}$ g cm$^{-3}$).
The heat is carried away by the neutrino emission
from the entire stellar body and by the thermal photon emission
from the surface. No additional reheating mechanisms are included.

The code calculates theoretical cooling curves, i.e., 
the effective surface temperature as detected by a distant observer,
$T_{\rm s}^\infty$, versus NS age $t$. The thermal history of 
an isolated NS consists
of three stages. The first is the stage of thermal relaxation
of the stellar interior (e.g., Lattimer et al.\ \cite{lattimeretal94},
Gnedin et al.\ \cite{gyp01}). It lasts for about 10--100 yr.
It is followed by the stage at which the NS interior is isothermal
and the neutrino luminosity exceeds the surface photon luminosity
($t \la (3-10)\times 10^5$ yr). 
The final stage is the photon cooling stage at which the photon
luminosity dominates the neutrino one.

In the NS crust we use the EOS of Negele \& Vautherin (\cite{nv73})
(atomic nuclei everywhere in the crust are assumed to be spherical).
The core-crust interface is placed
at the density $
1.5 \times 10^{14}$ g cm$^{-3}$.
A standard procedure is used to match the core
and crust EOSs near the core-crust interface.
In the core, we use two phenomenological EOSs
proposed by Prakash et al.\ (\cite{pal88}).
We refer to them as EOS A and EOS B. 

EOS A is model
I of Prakash et al.\ (\cite{pal88}) 
with the compression modulus of saturated
nuclear matter $K=240$ MeV. 
It has been used in Papers I and II.
EOS B corresponds to $K=180$ MeV
and to the simplified form of the symmetry energy
proposed by Page \& Applegate (\cite{pa92}).
EOS B has been used in a number
of papers (e.g., Page \& Applegate \cite{pa92}, Yakovlev 
et al.\ \cite{yls99}, and references therein).

The masses, central densities, and radii
of two stellar configurations for each EOS are given in Table 1.
The first configuration is the most massive
stable NS. The values of $M_{\rm max}$ indicate that EOS A
is stiff while EOS B is moderate.
The second configuration 
has a central density at which the 
direct Urca process switches on;
%
for both EOSs it is allowed
at $M>M_{\rm D}$
($\rho_{\rm c} > \rho_{\rm D}$). EOS B implies a smaller symmetry
energy at supranuclear densities and opens the direct Urca process
at a higher density.

\begin{table}[t]
\caption{NS models employing EOSs A and B}
\begin{tabular}{|l|l|l|l|}
\hline
Model  & Main parameters                       &     EOS A     &     EOS B     \\
\hline \hline
Maximum& $M_{\rm max}/{\rm M}_\odot$           &  1.977        &  1.73        \\
mass   & $\rho_{\rm cmax}/10^{14}$ g cm$^{-3}$ &  25.75        &  32.5         \\
model  & $R$ km                                &  10.754       &  9.71         \\
\hline \hline
Direct Urca& $M_{\rm D}/{\rm M}_\odot$             &   1.358       &  1.44     \\
threshold& $\rho_{\rm D}/10^{14}$ g cm$^{-3}$ &   7.851       &  12.98        \\
model              & $R$ km                                & 12.98& 11.54      \\
\hline
\end{tabular}
\end{table}

Our cooling code includes all the important neutrino emission
processes in the NS core (direct and modified Urca,
neutrino bremsstrahlung in nucleon-nucleon collisions,
neutrino emission due to Cooper pairing of nucleons) and 
in the crust (plasmon decay, neutrino bremsstrahlung due to
scattering of electrons off atomic nuclei, electron-positron 
annihilation into neutrino pairs, neutrino emission due
to Cooper pairing of free neutrons in the inner crust).
The effects of nucleon superfluidity are incorporated
in the neutrino reaction rates and nucleon heat capacities
as described in Yakovlev et al.\ (\cite{yls99,ykgh01}). 
The effective masses of protons and neutrons in the core and
free neutrons in the crust are taken to be 0.7 of the bare 
nucleon masses. The values of the thermal conductivity
in the NS crust and core are the same as used by Gnedin
et al.\ (\cite{gyp01}). 

The relationship between the effective surface temperature
and the temperature at the bottom of the outer heat-blanketing
envelope is taken according to Potekhin et al.\ (\cite{pcy97})
and Potekhin \& Yakovlev (\cite{py01}). This allows us to consider
either the models of NSs with the surface layers made of iron,
without magnetic field and with
the dipole surface magnetic fields $B \la 10^{15}$ G,
or the non-magnetic NS models with the surface layers
containing light elements. It is assumed that the
surface magnetic field induces an anisotropic heat transport
in the heat-blanketing envelope but does not violate 
the isotropic (radial) heat diffusion in the deeper NS layers.
It is also assumed that a NS may have a hydrogen atmosphere
even if the heat-blanketing envelope is mostly made of iron.
The majority of cooling curves will be calculated for 
the model of non-magnetized heat-blanketing envelope
made of iron. Two exceptions
are considered in Sect.\ 4.3.     

\begin{table}[t]
\caption{Parameters of superfluid models in Eq.\ (1)}
\begin{tabular}{lllllll}
Pair-     & Mo-   & $T_{0}/10^9~$K & $k_0$ &  $k_1$   & $k_2$    &    $k_3$ \\
ing       & del   &          & fm$^{-1}$  & fm$^{-1}$ & fm$^{-1}$& fm$^{-1}$ \\
\hline 
$^1$S$_0$ & 1p   & 20.29     &       0    &  1.117    & 1.241    &  0.1473  \\
$^1$S$_0$ & 2p   & 17        &       0    &  1.117    & 1.329    &  0.1179  \\
$^1$S$_0$ & 3p   & 14.5      &       0    &  1.117    & 1.518    &  0.1179  \\
$^1$S$_0$ & 1ns  & 10.2      &       0    &  0.6      & 1.45     &  0.1     \\
$^1$S$_0$ & 2ns  & 7.9       &       0    &  0.3      & 1.45     &  0.01 \\
$^1$S$_0$ & 3ns  & 1800      &       0    &   21      & 1.45     &  0.4125 \\
$^3$P$_2$ & 1nt  & 6.461     &       1    &  1.961    & 2.755    &  1.3
\end{tabular}
\end{table}

\begin{figure}
\centering
\epsfxsize=86mm
\epsffile[20 143 575 695]{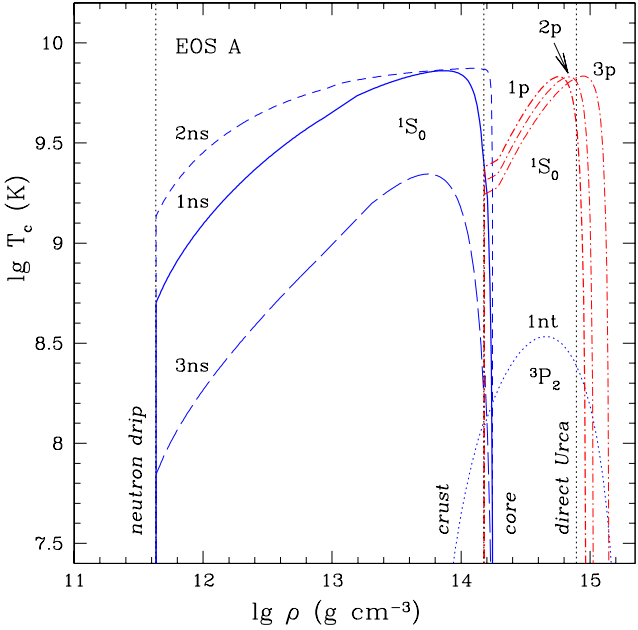}
\caption{
Density dependence of the critical temperatures
for three models 1p, 2p, and 3p
of the proton superfluidity (dots-and-dashes) in the core
(with EOS A); three models 
1ns, 2ns, and 3ns of $^1$S$_0$ neutron superfluidity 
(solid, short-dashed, and long-dashed lines);
and one model 1nt of $^3$P$_2$ neutron superfluidity (dots)
used in cooling
simulations. The parameters of the models are given in Table 2.
Vertical dotted lines indicate neutron drip point,
core-crust interface, and the direct Urca threshold.
}
\label{fig1}
\end{figure}

\begin{figure}
\centering
\epsfxsize=86mm
\epsffile[20 143 575 695]{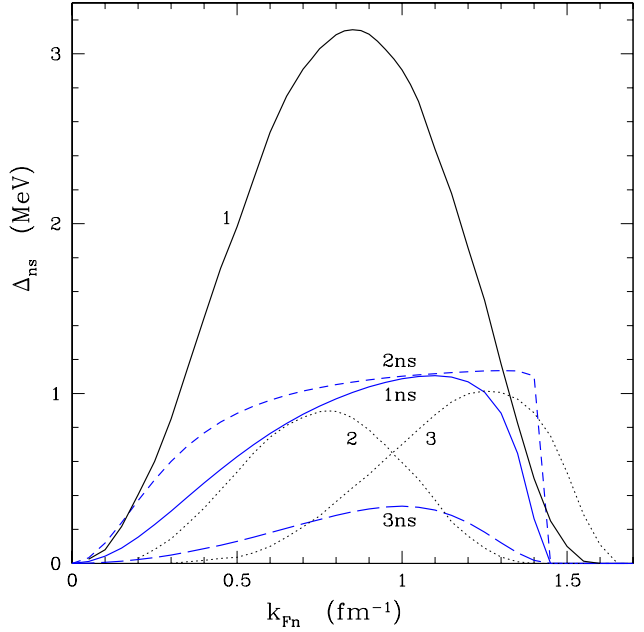}
\caption{
Superfluid gaps for $^1$S$_0$ neutron pairing
versus neutron Fermi wavenumber $k_{\rm Fn}$ for models 
1ns, 2ns, and 3ns
(solid line 1ns, short-dashed line 2ns, 
and long-dashed line 3ns; Fig. 1, Table 2).
Solid line 1 is the gap derived from the BCS theory
with in-vacuum nn interaction (after Lombardo \& Schulze \cite{ls01});
dotted lines 2 and 3 are two different theoretical curves obtained 
by Wambach et al.\  (\cite{wap93}) 
and Schulze et al.\ (\cite{schulzeetal96})
including medium polarization effects.
}
\label{fig2}
\end{figure}

\begin{figure}
\centering
\epsfxsize=86mm
\epsffile[20 143 575 695]{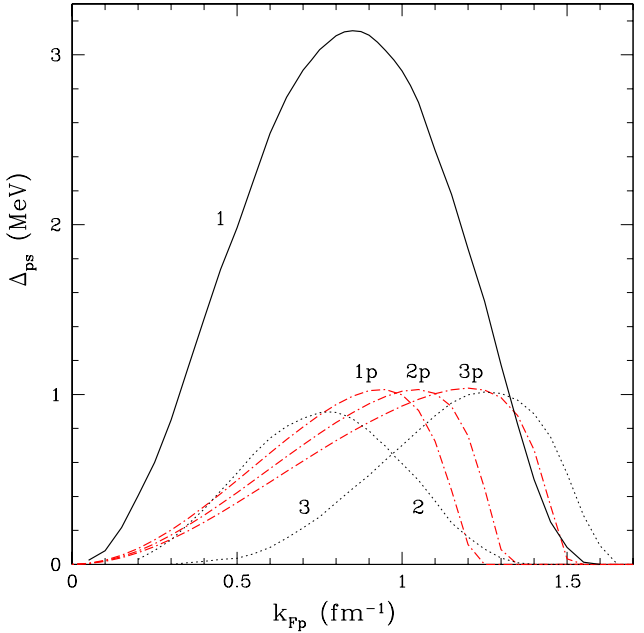}
\caption{
Superfluid gaps for models 1p, 2p, and 3p (Fig.\ 1, Table 2)
of proton pairing (dot-and-dashed lines) versus proton
Fermi wavenumber. Solid line 1 and dotted lines 2 and 3
are the same as in Fig.\ 2.
}
\label{fig3}
\end{figure}

\begin{figure}
\centering
\epsfxsize=86mm
\epsffile[30 30 350 410]{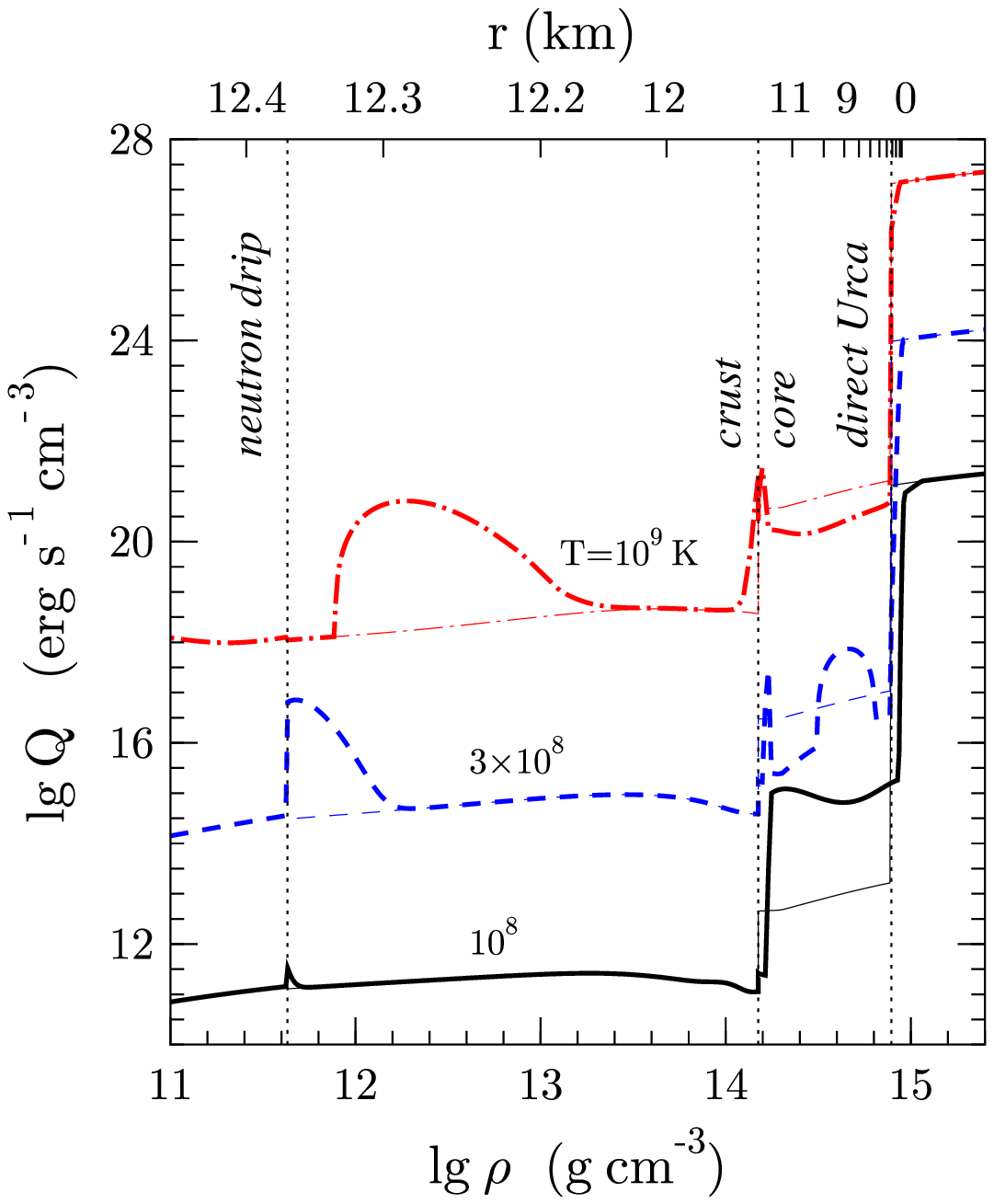}
\caption{
Total neutrino emissivity versus density (lower horizontal scale)
for $T=10^8$ (solid lines), $3 \times 10^8$ (dashes), and
$10^9$ K (dot-and-dashes) assuming EOS A at supranuclear densities. 
Thin lines are for non-superfluid matter;
thick lines are for matter with 1ns, 1nt, and 1p superfluids of
neutrons and protons.
Vertical dotted lines are the same as in Fig.\ 1.
Upper horizontal scale shows the radial coordinate $r$
in a 1.5 M$_\odot$ NS.
}
\label{fig4}
\end{figure}
 
Although the microscopic calculations of superfluid critical temperatures
$T_{\rm c}(\rho)$ give a large scatter of results (Sect.\ 1),
some common features are evident. For instance, $T_{\rm c}(\rho)$
increases with $\rho$ at sufficiently low densities  
(due to an increasing strength of the attractive 
part of nucleon-nucleon interaction), 
reaches maximum and then decreases (due to 
a short-range nucleon-nucleon repulsion)
vanishing at a rather high density. For $T_{\rm cns}(\rho)$,
the maximum takes place at subnuclear densities, while the switch off 
occurs at $\rho \sim \rho_0$, where $\rho_0 \approx 2.8 \times 10^{14}$
g cm$^{-3}$ is the saturated nuclear matter density. 
For $T_{\rm cp}(\rho)$ and $T_{\rm cnt}(\rho)$,
the maxima take place at a few $\rho_0$ and the fall occurs at
the densities several times higher. The maximum values of $T_{\rm c}$
range from about $10^8$ K (or even lower) to $(2 - 3) \times 10^{10}$ K,
depending on the microscopic theoretical model employed.
The maximum values of $T_{\rm cnt}$ are typically lower than
those of $T_{\rm cp}$ and $T_{\rm cns}$, due to the weaker nucleon-nucleon
attraction in the $^3$P$_2$ state.     

Instead of studying $T_{\rm c}$ as a function of $\rho$, it is often
convenient to consider 
it versus
the nucleon Fermi wavenumber
$k=k_{\rm F\!N}=(3 \pi^2 n_{\rm N})^{1/3}$, where $n_{\rm N}$
is the number density of nucleon species N=n or p. 
Moreover, instead of $T_{\rm c}(k)$ one often
considers $\Delta(k)$, the zero-temperature superfluid  
gap. For the $^1$S$_0$ pairing, assuming BCS theory,
one has $\Delta(k)=T_{\rm c}(k)/0.5669$.
In the case of $^3$P$_2$ neutron pairing
the gap depends on the orientation of nucleon momenta with
respect to the quantization axis. 
Following the majority of papers we adopt the $^3$P$_2$
pairing with zero projection of 
the total angular momentum on the quantization
axis. In that case $\Delta_{\rm nt}(k)=T_{\rm cnt}(k)/0.8416$,
where $\Delta_{\rm nt}(k)$ is the minimum value of the gap
on the neutron Fermi surface 
(corresponding to
the equator of the Fermi
sphere; e.g., Yakovlev et al.\ \cite{yls99}).
  
Taking into account a large scatter of 
$T_{\rm c}(k)$ provided by microscopic theories
we do not rely on any particular microscopic model. 
Following Papers I and II, we parameterize $T_{\rm c}$ as 
\begin{equation}
    T_{\rm c}= T_0 \, {(k-k_0)^2 \over (k-k_0)^2 + k_1^2} 
     \; {(k-k_2)^2 \over  (k-k_2)^2 + k_3^2 }~,
\label{Tc}
\end{equation}
for $k_0< k < k_2$; and $T_{\rm c}=0$, for $k \leq k_0$ or $k \geq k_2$.
The factor $T_0$ regulates the amplitude of $T_{\rm c}$, $k_0$ and $k_2$ determine
positions of the low- and high-density cutoffs,
while $k_1$ and $k_3$ specify the shape of $T_{\rm c}(\rho)$.
All wavenumbers, $k$, $k_0$, \ldots $k_3$ will be expressed
in fm$^{-1}$. 
We have verified that by tuning $T_0$, $k_0$, \ldots  $k_3$,
our parameterization describes accurately
numerous results of microscopic calculations.

In our cooling simulations
we consider three models of $^1$S$_0$ proton superfluidity,
three models of $^1$S$_0$ neutron superfluidity,
and one model of $^3$P$_2$ neutron superfluidity.
The parameters of the models are given in Table 2,
and the appropriate $T_{\rm c}(\rho)$ are plotted in Fig.\ \ref{fig1}.
We have $k_0=0$ for $^1$S$_0$ pairing. At any given $\rho$
we choose the neutron superfluidity ($^1$S$_0$ or $^3$P$_2$)
with higher $T_{\rm c}$.

Models 1ns, 2ns, and 3ns of $^1$S$_0$ neutron superfluidity
are the same as used in Paper II.
Models 1ns and 2ns correspond to about the same, rather strong 
superfluidity (with maximum $T_{\rm cns} \approx 7 \times 10^9$ K). 
Model 2ns has flatter maximum 
and sharper decreasing slopes in the wings
(near the crust-core interface and the neutron drip point). Model 3ns 
represents a much weaker superfluidity, with maximum 
$T_{\rm cns}\approx 2.4 \times 10^9$ K and a narrower 
density profile. (One can visualize the radial
distributions of $T_{\rm c}$ in a NS by comparing
the horizontal scales in Figs.\  \ref{fig1} and \ref{fig4}.)  
 
The superfluid gaps for these models are shown in Fig.\ \ref{fig2}
versus the neutron Fermi wavenumber.
For comparison, we present also three curves provided 
by microscopic theories. Solid curve 1 is obtained using BCS theory
with the in-vacuum nn-interaction 
(after Lombardo \& Schulze \cite{ls01}). 
This approach yields a very strong
superfluidity with the maximum gap $\Delta_{\rm ns} \approx 3$ MeV.
Dotted curves 2 and 3 are calculated using two different models
of nn-interaction affected by the medium polarization 
(Wambach et al.\ \cite{wap93}, Schulze et al.\ \cite{schulzeetal96}).
The polarization effects strongly reduce the gaps.
Microscopic models of $\Delta_{\rm ns}(k)$ are abundant in the literature and the results
differ considerably 
(see, e.g., Fig.\ 7 in Lombardo \& Schulze \cite{ls01}
or Fig.\ 3 in Yakovlev et al.\ \cite{yls99}). 
Our phenomenological curves 1ns, 2ns, and 3ns 
all fall in the range covered by theoretical curves.
The shapes of the $\Delta_{\rm ns1}(k)$ 
and $\Delta_{\rm ns3}(k)$ curves 
are typical 
whereas the decrease of $\Delta_{\rm ns2}(k)$
with $k$ at $k \ga 1.4$ fm$^{-1}$ 
is not (too sharp).
 
The proton superfluidity curves 1p, 2p, and 3p in Fig.\ 1 are similar.
The maximum values of $T_{\rm cp}$ are about $7 \times 10^9$ K for all
three models. Note that model 1p was used in Papers I and II.
The models differ by the positions of the maximum and decreasing
slopes of $T_{\rm cp}(\rho)$.
The decreasing slope of model 1p 
is slightly above the threshold density
of the direct Urca process (for EOS A), while the slopes for models
2p and 3p are shifted to a higher $\rho$.
The corresponding gaps   
are shown in Fig.\ \ref{fig3}. 
For comparison, in Fig.\ \ref{fig3} 
we present the same curves 1, 2, and 3 as
in Fig.\ \ref{fig2} (the proton gap $\Delta_{\rm ps}(k_{\rm Fp})$
is expected to be similar to the neutron gap 
$\Delta_{\rm ns}(k_{\rm Fn})$).
Our models 1p, 2p, and 3p are typical for those 
microscopic theories which adopt a moderately 
strong medium polarization of
pp-interaction. 

Finally, the dotted curve in Fig.\ \ref{fig1} 
shows $T_{\rm cnt}(\rho)$
for our model 1nt of $^3$P$_2$ neutron pairing
(used in Paper I).
Microscopic theories give a very large scatter of $T_{\rm cnt}(\rho)$,
and our curve falls within their limits.  

\newcommand{\rrr}{\rule{0cm}{0.4cm}}
\newcommand{\hh}{\rule{0.5cm}{0cm}}
\newcommand{\hb}{\rule{0.4cm}{0cm}}
 
\begin{table*}[!t]   
\caption[]{Surface temperatures of eight isolated middle-aged neutron stars
inferred from observations}
\label{tab-cool-data}
\begin{center}
\begin{tabular}{|| l | l | l | l | c | l ||}
\hline
\hline
 Source & lg~t & lg~$T_{\rm s}^\infty$ & Model$^{a)}$ & Confid. & References   \\
              & [yr] &  [K]             &              & level   &           \\ 
\hline
\hline
RX$\,$J0822--43 & 3.57 & $ ~~~6.23^{+0.02 \rrr}_{-0.02} $ & H  &  95.5\% &
Zavlin et al.\ (\cite{ztp99}) \\
1E$\,$1207--52  & 3.85 & $ ~~~6.10^{+0.05 \rrr}_{-0.06} $  & H  & 90\% &
Zavlin et al.\ (\cite{zpt98}) \\
RX$\,$J0002+62 & $3.95^{b)}$ & $ ~~~6.03^{+0.03 \rrr}_{-0.03} $ & H & 95.5\%&
Zavlin \& Pavlov (\cite{zp99}) \\
PSR~0833--45 (Vela) & $4.4^{c)}$ & $ ~~~5.83^{+0.02 \rrr}_{-0.02}$ & H & 68\%  &
Pavlov et al.\ (\cite{pavlovetal01})\\
PSR~0656+14 & 5.00 & $ ~~~5.96^{+0.02\rrr}_{-0.03} $ & bb &  90\%  &
Possenti et al.\ (\cite{pmc96}) \\
PSR~0633+1748 (Geminga) & 5.53 & $ ~~~5.75^{+0.05\rrr}_{-0.08} $ & bb & 90\% &
Halpern \& Wang (\cite{hw97}) \\
PSR~1055--52 & 5.73 & $ ~~~5.88^{+0.03\rrr}_{-0.04} $ & bb & $^{d)}$ &
\"{O}gelman (\cite{ogelman95}) \\
RX~J1856--3754 & 5.95 & $ ~~~5.72^{+0.05\rrr}_{-0.06} $ 
& $^{e)}$  & $^{d)}$ &
Pons et al.\ (\cite{ponsetal01}) \\
\hline
\end{tabular}
\begin{tabular}{l}
  $^{a)}\,$\rrr{\footnotesize Observations are interpreted either with 
a hydrogen atmosphere model (H), or with a black body spectrum (bb)}\\[0.5ex]
  $^{b)}\,$\rrr{\footnotesize The mean age taken according
 to Craig et al.\ (\cite{chp97}).}\\[-0.5ex]
  $^{c )}\,$\rrr{\footnotesize According to
        Lyne et al.\ (\cite{lyneetal96}).}\\[-0.5ex]
  $^{d)}\,$\rrr{\footnotesize Confidence level is uncertain.}\\[0.5ex] 
  $^{e)}\,$\rrr{\footnotesize Analytic fit with Si-ash atmosphere model 
     of Pons et al.\ (\cite{ponsetal01}).} 
%
\end{tabular}
\end{center}
\end{table*}

Figure 4 shows the density profile of the neutrino emissivity
at $T=10^8$, $3 \times 10^8$, and $10^9$ K. Thin lines
correspond to non-superfluid matter and have three distinct
parts. In the crust, the emissivity is mainly produced
by neutrino bremsstrahlung due to the scattering of electrons
off nuclei. In the outer core, the emissivity is 
mainly produced by the modified Urca process and is
about two orders of magnitude higher.
In the inner core, it is due to the direct Urca process and
is higher by another 6--7 magnitudes.
Thick lines are for superfluid matter
(1ns, 1nt, and 1p superfluids of neutrons and protons). 
At $T=10^9$ K there are two
large emissivity peaks, near the neutron drip-point and the crust-core
interface. They are associated with the neutrino emission
due to $^1$S$_0$ Cooper pairing of neutrons. 
They are explained by the fact 
that the Cooper-pairing neutrino emissivity 
is most intense at $0.8 \, T_{\rm c} \la T \leq T_{\rm c}$,
and is exponentially small at $T \ll T_{\rm c}$ 
(e.g., Yakovlev et al.\ \cite{yls99}). 
One can also see a reduction
of neutrino emission in the outer core 
by the proton superfluidity. The same
superfluidity reduces also the direct Urca process near
its threshold, $\rho \ga \rho_{\rm D}$, but becomes weaker
(Fig.\ \ref{fig1}) and has no effect at higher densities.
At $T=3 \times 10^8$ K the peaks
associated with $^1$S$_0$ neutron pairing are weaker,
but there is a new high peak in the outer core due
to $^3$P$_2$ neutron pairing. At
$T=10^8$ K the neutrino emission due to 
$^1$S$_0$ neutron pairing almost disappears but
the emission due to $^3$P$_2$ pairing 
persists.  At still lower temperature, the
$^3$P$_2$-pairing emissivity in the outer core will have two peaks
and gradually disappear. Upper horizontal scale
gives the radial coordinate in a 1.5 M$_\odot$ NS.
The bulk of neutrino emission comes evidently from the
core ($\sim 11$ km in radius), 
and a lower fraction comes from the crust
($\sim 1$ km thick). 

The results of cooling calculations are illustrated
in Figs.\ \ref{fig5}-\ref{fig10} and described in Sect.\ 4.
 
\section{Observational data}
\label{sect3}

We will confront theoretical cooling curves with the results
of observations of thermal emission from eight middle-aged isolated
NSs. The observational data are the same as in Papers I and II. 
They are summarized in Table 3 and displayed
in Figs.\ 5, 6, 7, and 10. 
The three youngest objects 
(RX J0822--43, 1E 1207--52, and RX J0002+62) are radio-quiet
NSs in supernova remnants. The oldest object, RX J1856--3754,
is also a radio-quiet NS. The other objects, Vela, PSR 0656+14,
Geminga, and PSR 1055--52, are observed as radio pulsars.
The NS ages are either pulsar spindown ages or the estimated
supernova ages. The age of RX J1856--3754 was estimated
by Walter (\cite{walter01}) from the kinematical data
(by identifying  a possible presupernova companion
in the binary system).
We use the value $t=9 \times 10^5$ yr
mentioned in the subsequent publication by Pons et al.\ 
(\cite{ponsetal01}).

For the four youngest sources, the effective surface temperatures 
$T_{\rm s}^\infty$
are obtained from the observed X-ray spectra using
hydrogen atmosphere models. Such models are more consistent with other
information on these sources (distances, hydrogen column
densities, inferred NS radii, etc.) than the blackbody model of
NS emission. On the contrary, for the next three sources we present
the values of $T_{\rm s}^\infty$ inferred using the blackbody spectrum
because the blackbody model is more consistent for these sources.
Finally, for RX J1856--3754 we adopt the values inferred
using the analytic fit with Si-ash atmosphere model of Pons et al.\
(\cite{ponsetal01}). We expect that the large errorbar of $T_{\rm s}^\infty$
provided by this model reflects poor understanding
of thermal emission from this source
(e.g., Pons et al.\ \cite{ponsetal01}, Burwitz et al.\ \cite{burwitzetal01},
G\"ansicke et al.\ \cite{gbr01}, Kaplan et al.\ \cite{kva01}).

\section{Theory versus observations}
\label{sect4}

\subsection{General remarks}
\label{sect4-1}

We have a large scatter of observational limits on $T_{\rm s}^\infty$
for the eight sources. Three sources, the youngest RX J0822--43,
and two oldest, PSR 1055--52 and RX J1856--3754,  seem to be hot for their
ages, while the other ones, especially Vela and Geminga, look much colder.
Our aim is to interpret all observational data 
with the cooling curves using the fixed (the same) EOS and
models of the critical temperatures $T_{\rm c}(\rho)$ (Sect.\ 2)
for all objects.    
The results are presented in
Figs.\ \ref{fig5}--\ref{fig10}.

In the {\it absence of any superfluidity} 
we would have {\it two} well-known, distinctly different
cooling regimes, {\it slow} and {\it fast} cooling.
The slow cooling takes place in low-mass
NSs, with $M< M_{\rm D}$. In middle-aged
NSs, it goes mainly via neutrino emission produced by modified
Urca processes. For a given EOS in the NS core, the cooling
curves of middle-aged NSs are almost the same 
for all $M$ from about $1.1 \, {\rm M}_\odot$
to $M_{\rm D}$ (e.g., Page \& Applegate \cite{pa92},
Gnedin et al.\ \cite{gyp01}), and they are not very
sensitive to EOS.
The fast cooling occurs
if $M > M_{\rm D}+ 0.003 \, {\rm M}_\odot$
via a very powerful direct Urca process
(Lattimer et al.\ \cite{lpph91}).
The cooling curves are again not too sensitive to the mass and EOS.
The middle-aged rapidly cooling NSs are much colder
than the slowly cooling ones.
Two examples, for 1.35 and 1.5 M$_\odot$ 
non-superfluid NSs (EOS A), 
are displayed
in Fig.\ \ref{fig5} by long dashes.
The transition from the slow to fast cooling takes
place in a very narrow range of $M$. It
is demonstrated in Fig.\ \ref{fig8} which shows
$T_{\rm s}^\infty$ versus NS mass at the
age $t=2.5 \times 10^4$ yr of the Vela
pulsar. Horizontal dotted lines show
observational limits on $T_{\rm s}^\infty$ for the Vela pulsar.
One can see a very sharp fall of $T_{\rm s}^\infty$
in the mass range
$M_{\rm D}< M \la M_{\rm D}+0.003 \, {\rm M}_\odot$ 
for non-superfluid NSs followed by a slow fall at 
$M \ga  M_{\rm D}+0.003 \, {\rm M}_\odot$.
In order to explain these observational limits with
the non-superfluid NS models we should make an unlikely assumption
that the Vela's mass falls in that narrow mass range.
We would face the same difficulty with 
1E 1207--52, RX J0002+62, PSR 0656+14, and Geminga.
Thus we have five sources which exhibit
the intermediate case between the slow and fast cooling.
In the absence of superfluidity, this is highly unlikely.

\subsection{Proton superfluidity and the three types of cooling neutron stars}
\label{sect4-2}

\begin{figure}
\centering
\epsfxsize=86mm
\epsffile[20 143 590 720]{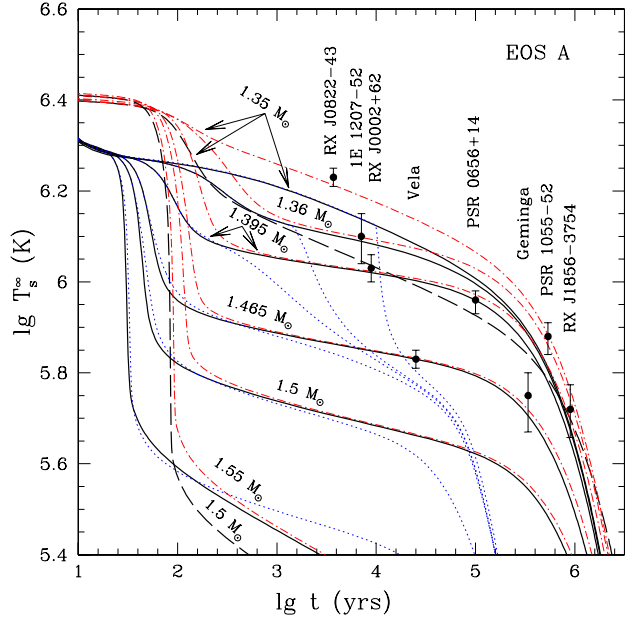}
\caption{
Observational limits on surface temperatures of eight
NSs (Table 3) compared with cooling curves
for NS models (EOS A) with masses from
1.35 to 1.55 ${\rm M}_\odot$.
Dot-and-dashed curves are obtained
including proton superfluidity 1p alone.
Solid curves include, in addition,
model 1ns of neutron superfluidity.
Dotted lines also take into account
the effect of neutron superfluidity 1nt.
Long-dashed lines are for non-superfluid 
1.35 and 1.5 M$_\odot$ NSs. 
}
\label{fig5}
\end{figure}

\begin{figure}
\centering
\epsfxsize=86mm
\epsffile[20 143 590 720]{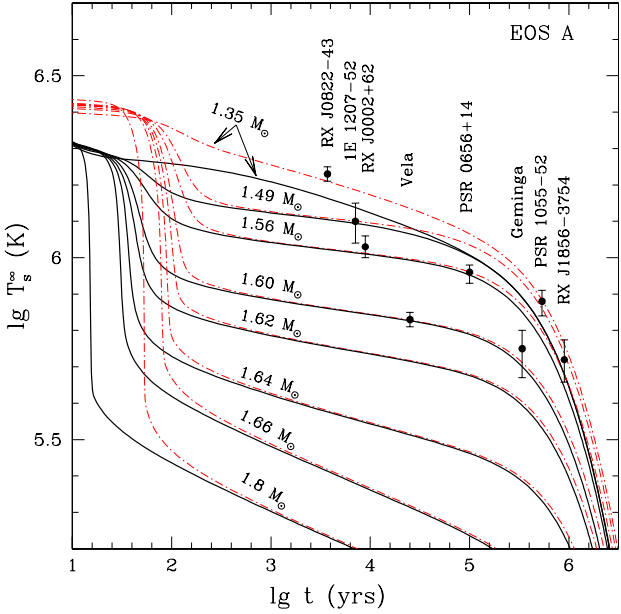}
\caption{
Observational limits on surface temperatures of 
NSs compared with cooling curves
for NS models (EOS A) with several masses $M$ in the presence of
proton superfluidity 2p.
Dot-and-dashed curves are obtained
assuming non-superfluid neutrons.
Solid curves include, in addition,
model 1ns of neutron superfluidity.
$^3$P$_2$ neutron pairing is neglected.
}
\label{fig6}
\end{figure}

\begin{figure}
\centering
\epsfxsize=86mm
\epsffile[20 143 590 720]{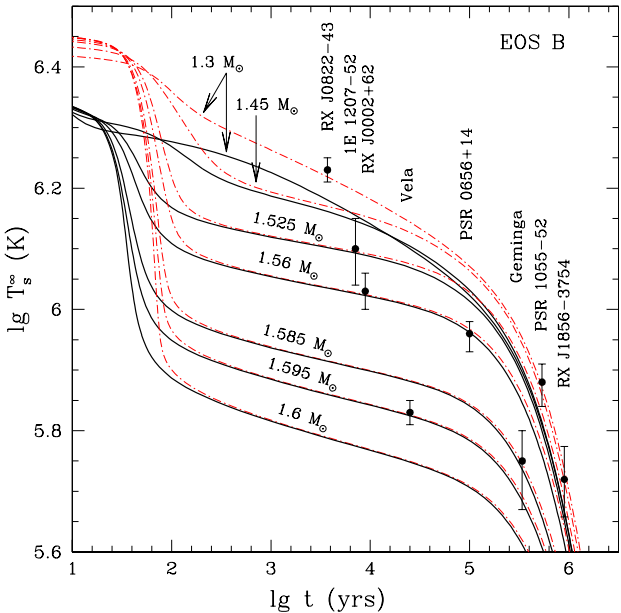}
\caption{
Observational limits on surface temperatures of 
NSs compared with cooling curves
for NS models (EOS B) with masses from
1.3 to 1.6 ${\rm M}_\odot$.
Dot-and-dashed curves are obtained
using model 3p of proton superfluidity.
Solid curves include, in addition,
model 1ns of neutron superfluidity.
}
\label{fig7}
\end{figure}

Let us explain the observations 
(Fig.\ \ref{fig5}--\ref{fig7}) by cooling of superfluid NSs. 
It turns out (Papers I and II) that various superfluids
affect NS cooling in different ways. 
Our main assumptions would be that 
(i) the {\it proton superfluidity
is rather strong} at $\rho \la \rho_{\rm D}$, while 
(ii) the $^3$P$_2$
{\it neutron superfluidity is rather weak} (Sect.\ 4.4).
We will discuss the superfluid effects
step by step starting from the effects of proton superfluidity.

The dot-and-dashed cooling curves in Figs.\ \ref{fig5}--\ref{fig7}
are computed
assuming the proton superfluidity alone. We adopt the proton pairing 1p
in Fig.\ \ref{fig5}, 2p in Fig.\ \ref{fig6}, 
and 3p in Fig.\ \ref{fig7}. We use EOS A in the models
in Figs.\ \ref{fig5} and \ref{fig6}, and EOS B in Fig.\ \ref{fig7}.

\begin{figure}
\centering
\epsfxsize=86mm
\epsffile[20 143 590 720]{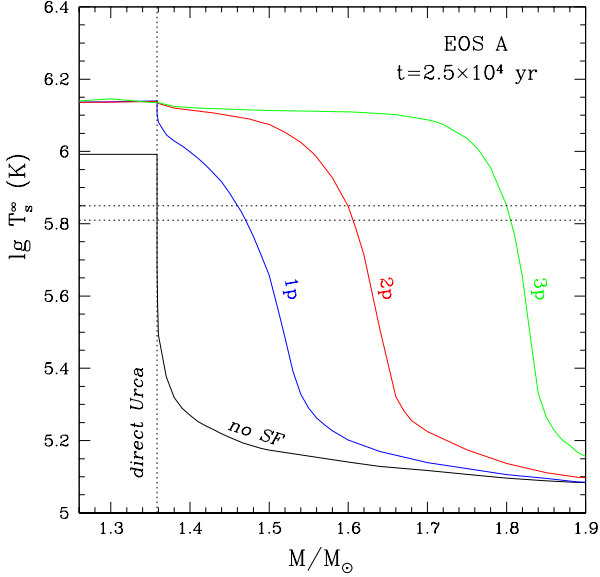}
\caption{
Surface temperatures of NS models (EOS A)
at $t=2.5 \times 10^4$ yr, the age of the Vela
pulsar, versus stellar mass $M$.
The curves are calculated (from bottom to top)
assuming either no superfluidity, or proton superfluidities 1p, 2p, 3p
(and neglecting neutron superfluidity).
The vertical dotted line shows the threshold mass $M_{\rm D}$
of opening the direct Urca process. Horizontal dotted lines
are observational limits on the Vela's surface temperature 
(Table 3).
}
\label{fig8}
\end{figure}

Analyzing Figs.\ \ref{fig5}--\ref{fig8} we 
see that, generally, the proton superfluidity
leads to the {\it three} cooling regimes
(instead of two): {\it slow, moderate}, and {\it fast}. 
Accordingly, we predict
{\it three types} of cooling NSs with distinctly
different properties.

(I) {\it Low-mass, slowly cooling} NSs.
The central densities $\rho_{\rm c}$ 
and masses $M=M(\rho_{\rm c})$ of these NSs obey
the inequalities
\begin{equation}
  \rho_{\rm c}\la \rho_{\rm I}, \quad      
  M \la M_{\rm I} = M(\rho_{\rm I});
\label{MI}
\end{equation}
the threshold values 
$\rho_{\rm I}$ and $M_{\rm I}$ are specified below.

(II) {\it Medium-mass, moderately cooling} NSs, with
\begin{equation}    
\rho_{\rm I} \la \rho_{\rm c} \la \rho_{\rm II},
\quad \quad    
M_{\rm I} \la M \la M_{\rm II}=M(\rho_{\rm II}),
\label{MII}
\end{equation}
$\rho_{\rm II}$ and $M_{\rm II}$ are also
specified below.

(III) {\it Massive, rapidly cooling} NSs,
\begin{equation}
  \rho_{\rm II} \la \rho_{\rm c} \leq \rho_{\rm cmax}, \quad 
   M_{\rm II} \la M \leq M_{\rm max},
\label{MIII}
\end{equation}
where $\rho_{\rm cmax}$ and $M_{\rm max}$ refer to
the maximum-mass configuration (Table 1). 

We will show that the threshold values of $\rho_{\rm I,II}$ and
$M_{\rm I,II}$ depend on a proton superfluidity model,
EOS in the NS core, and on 
a NS age. Let us describe these three cooling regimes
in more detail.

(I) We {\it define the slowly cooling NSs} 
as those where the direct Urca
process is either forbidden by momentum conservation
($\rho_{\rm c} \leq \rho_{\rm D}$, Lattimer et al.\ \cite{lpph91})
or greatly suppressed by the strong proton superfluidity. 

In particular, we have the 
slow cooling for $\rho_{\rm c} < \rho_{\rm D}$
and $M < M_{\rm D}$ in the absence of proton superfluidity.
This is the {\it ordinary} slow cooling discussed 
widely in the literature. It is mainly regulated by
the neutrino energy losses produced by the
modified Urca process.

However, for the conditions displayed in 
Figs.\ \ref{fig5}--\ref{fig7}, 
the proton superfluidity is so strong
that it almost switches off both,
the modified Urca process everywhere in the NS core
and the direct Urca process at $\rho > \rho_{\rm D}$.
Then the main neutrino emission
is produced by {\it neutrino bremsstrahlung in
neutron-neutron collisions} (unaffected by
the neutron superfluidity in the NS core
that is assumed to be weak).
The bremsstrahlung is less efficient than
the modified Urca process and leads
to an even slower cooling than in a non-superfluid NS. We will refer to it
as the {\it very slow} cooling. 

The analysis shows that,
for our proton superfluid models,
the regime of very slow cooling holds as long as
the proton critical temperature in the NS center is higher
than a threshold value: 
\begin{equation}
      T_{\rm cp}(\rho_{\rm c}) \ga T_{\rm cp}^{\rm (I)}(\rho_{\rm c}).
\label{TcpI}
\end{equation}
Comparing the neutrino emissivities of the indicated
reactions (e.g., Yakovlev et al.\ \cite{yls99}), we
can obtain the simple estimates:
$T_{\rm cp}^{\rm (I)}(\rho) \sim 5.5 \, T$ for 
$\rho \leq \rho_{\rm D}$, and
$T_{\rm cp}^{\rm (I)}(\rho) \sim 17 \, T$ for
$\rho$ several per cent higher than $\rho_{\rm D}$,
where $T$ is the internal NS temperature.
There is a continuous transition
between the presented values of $T_{\rm cp}^{\rm (I)}(\rho)$
in the narrow density range near $\rho_{\rm D}$.

\begin{figure}
\centering
\epsfxsize=86mm
\epsffile[20 143 590 720]{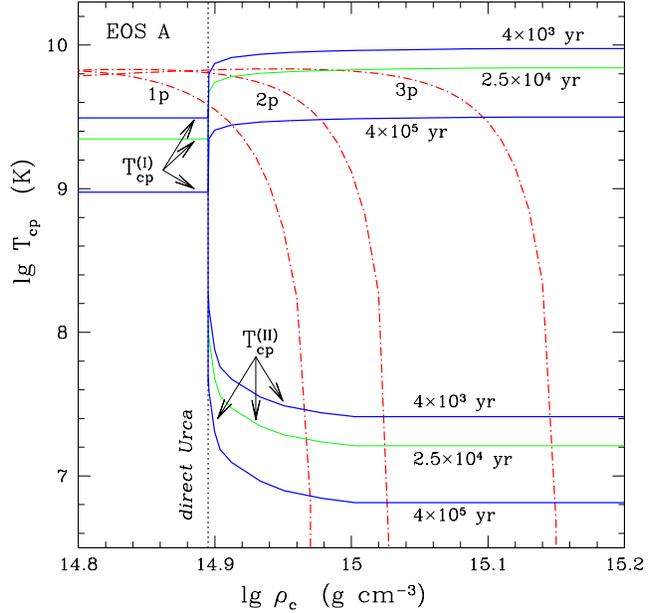}
\caption{
Two characteristic critical
temperatures of proton superfluidity, $T_{\rm cp}^{\rm (I)}$
and $T_{\rm cp}^{\rm (II)}$, 
versus NS central density $\rho_{\rm c}$ for EOS A
and three NS ages, $4 \times 10^3$, $2.5 \times 10^4$, and
$4 \times 10^5$ yr. Also shown are $T_{\rm cp}(\rho)$
(dot-and-dashed) curves for proton superfluids 1p, 2p, and 3p. Their
intersections with $T_{\rm cp}^{\rm (I,II)}$ at
$\rho_{\rm c} \geq \rho_{\rm D}$
separates the low-, medium-, and high-mass NS models (see
text for details).
}
\label{fig9}
\end{figure}

To make our analysis less abstract we notice that
$T \sim 5.5 \times 10^8$ K in a very slowly cooling NS
at $t \sim 4 \times 10^3$ yr,
$T \sim 4 \times 10^8$ K at $t \sim 2.5 \times 10^4$ yr,
and $T \sim 1.5 \times 10^8$ K at $t \sim 4 \times 10^5$ yr.
The dependence of $T_{\rm cp}^{\rm (I)}$ on
$\rho_{\rm c}$ and on the
NS age is shown in Fig.\ \ref{fig9}. 

Now we are ready to specify the maximum 
central densities $\rho_{\rm I}$ and masses $M_{\rm I}$
of slowly cooling NSs in Eq.\ (\ref{MI}). 
For the cases of study, we have $\rho_{\rm I} \geq \rho_{\rm D}$ and
$M_{\rm I} \geq M_{\rm D}$ because the NSs with
$\rho_{\rm c}< \rho_{\rm D}$ show slow cooling.  
If $\rho_{\rm c} > \rho_{\rm D}$, then
we may find the density $\rho_{\rm I}$
on the decreasing, high-density slope of
$T_{\rm cp}(\rho)$ (Fig.\ \ref{fig1})
which corresponds to 
$T_{\rm cp}(\rho_{\rm I}) = 
T_{\rm cp}^{\rm (I)}(\rho_{\rm I})$ (Fig.\ \ref{fig9}).
It gives the central density of the star 
and $M_{\rm I}=M(\rho_{\rm I})$. If  $\rho_{\rm I}$ is formally
lower than $\rho_{\rm D}$, we set $M_{\rm I}=M_{\rm D}$.

Table 4 shows the values of $M_{\rm I}$ for EOSs A and B and
proton superfluids 1p, 2p, and 3p at two NS ages,
$t_1=4 \times 10^3$ yr and $t_2=4 \times 10^5$ yr.

According to Fig.\ \ref{fig8}, a sufficiently strong
proton superfluidity smears out a sharp transition from
the slow to fast cooling as the mass grows.
This effect is illustrated for all three proton 
superfluid models and EOS A at $t=2.5 \times 10^4$ yr. 
In models 2p and 3p
the proton superfluidity at the direct Urca threshold
is very strong. It drastically
suppresses the  
direct Urca process
and makes it unimportant. In these cases, the direct
Urca threshold {\it does not manifest a transition to
faster cooling}. Thus, at $t=2.5 \times 10^4$ yr
we have $\rho_{\rm I} \sim \rho_{\rm D}$
and $M_{\rm I} \sim M_{\rm D}$ for proton superfluid
1p, and we have $\rho_{\rm I} > \rho_{\rm D}$, $M_{\rm I} > M_{\rm D}$
for superfluids 2p and 3p (Fig.\ \ref{fig9}). 

\begin{table}[t]
\caption{Masses $M_{\rm I,II}$ which separate
slow, moderate, and rapid cooling models for $t_1=4 \times 10^3$
and $t_2=4 \times 10^5$ yr}
\begin{tabular}{ll|ll|ll}
EOS   & Proton &   
\multicolumn{2}{c|}{$  M_{\rm I}/{\rm M}_\odot$} 
& 
\multicolumn{2}{c}{$  M_{\rm II}/{\rm M}_\odot$}
\\
      & pairing &  $t_1$     & $t_2$         & $t_1$     & $t_2$  \\
\hline
A     & 1p   & $M_{\rm D}$   & 1.4           & 1.52      & 1.53   \\
A     & 2p   & $M_{\rm D}$   & 1.55          & 1.64      & 1.64   \\
A     & 3p   & $M_{\rm D}$   & 1.77          & 1.83      & 1.84   \\
B     & 1p   & $M_{\rm D}$   &  $M_{\rm D}$  & $M_{\rm D}$& $M_{\rm D}$  \\
B     & 2p   & $M_{\rm D}$   &  $M_{\rm D}$  & $M_{\rm D}$& $M_{\rm D}$ \\
B     & 3p   & $M_{\rm D}$   & 1.55          & 1.62      & 1.62     
\end{tabular}
\end{table}

For the conditions displayed 
in Figs.\ \ref{fig5}--\ref{fig7}, the
cooling curves (dot-and-dashed lines) of all
low-mass NSs are very similar.
For instance,
the $1.35 \, {\rm M}_\odot$ curve in Fig.\ \ref{fig5}
is plotted just as an example; all the curves are almost 
identical in the mass range $1.1 \, {\rm M}_\odot \la M < M_{\rm I}$.
Moreover, the curves are not too sensitive to EOS
and are {\it insensitive to the exact values of the
proton critical
temperature} $T_{\rm cp}$,
as long as the inequality (\ref{TcpI}) holds.
They are noticeably higher than the analogous cooling curves 
of the ordinary slow cooling in the
absence of superfluidity (e.g., Fig.\ \ref{fig5}).

For the conditions in Figs.\ \ref{fig5}--\ref{fig7}
(as in Papers I and II) we may explain
the three relatively hot sources,
RX J0822--43, PSR 1055--52, and RX J1856--3754, by
these  very-slow-cooling 
models with a strong proton superfluidity. 
Thus, we assume that the indicated sources are
low-mass NSs.
We discuss this explanation further in Sect.\ 4.3.

(II) We {\it define the moderately cooling stars} 
as the NSs which possess central kernels
where the direct Urca process is allowed but
moderately suppressed by proton superfluidity.
The existence of a representative class of these NSs
is solely due to proton superfluidity.

Our analysis shows that, for our cooling models,
the proton critical temperature 
in the center of a medium-mass NS should roughly satisfy the inequality
\begin{equation}
   T_{\rm cp}^{\rm (II)}(\rho_{\rm c}) 
   \la T_{\rm cp}(\rho_{\rm c}) \la
   T_{\rm cp}^{\rm (I)}(\rho_{\rm c}), 
\label{TcpII}
\end{equation}
with $T_{\rm cp}^{\rm (II)} \sim 3 \, T$.
Thus, we may introduce $\rho_{\rm II}$ which corresponds
to $T_{\rm cp}(\rho_{\rm II}) = T_{\rm cp}^{\rm (II)}(\rho_{\rm II})$
and determines
$M_{\rm II}$, the maximum mass of moderately
cooling NSs in Eq.\ (\ref{MII}). The values of
$T_{\rm cp}^{\rm (II)}$ depend mainly
on a given EOS and slightly
on a NS age (Fig.\ \ref{fig9}). The ranges of mass
and density in Eq.\ (\ref{MII}) depend
also on a model of $T_{\rm cp}(\rho)$.

The values of $M_{\rm II}$ are also given in Table 4,
along with $M_{\rm I}$.
For EOS B, the critical temperature $T_{\rm cp}(\rho)$
of 1p and 2p proton superfluids vanishes at $\rho \la \rho_{\rm D}$.
Then we have a sharp transition from the slow to fast
cooling in a narrow mass range just as
in the absence of the superfluidity (Sect.\ 4.1;
$M_{\rm I} \approx M_{\rm II} \approx M_{\rm D}$), and the regime
of moderate cooling is almost absent. 

The surface temperatures of the medium-mass 
(moderately cooling) NSs
are governed by proton superfluidity in the NS
central kernels, $\rho \ga \rho_{\rm I}$. 
One can observe (Figs.\ \ref{fig5}--\ref{fig8})
a steady decrease of   
surface temperatures with increasing $M$.
If we fix the proton superfluidity and EOS
(provided they allow for the moderate cooling)
we can determine (Papers I and II) the mass of any 
moderately cooling NS, which means
``weighing'' NSs.
In this fashion we can weigh five isolated NSs
(1E 1207--52, RX J0002+62, Vela, PSR 0656+14, and
Geminga) using either EOS A and the proton superfluids 1p, 2p, and 3p,
or EOS B and the superfluid 3p.
Thus, we assume that the indicated sources are moderately
cooling NSs. For instance, adopting
EOS A and proton superfluid 1p (Fig.\ \ref{fig5}) 
we obtain the masses in the range from
$\approx 1.36 \, {\rm M}_\odot$ (for 1E 1207--52) to 
$\approx 1.465 \, {\rm M}_\odot$ (for Vela and Geminga).
For EOS A and proton superfluid 2p (Fig.\ \ref{fig6})
we naturally obtain higher masses of the same sources.
Obviously, the properties of moderately cooling NSs
are {\it extremely sensitive} to the decreasing
slope of $T_{\rm cp}(\rho)$ in the temperature
range from $T_{\rm cp}^{\rm (II)}$ to $T_{\rm cp}^{\rm (I)}$
(Fig.\ \ref{fig9}), or in the density range from $\rho_{\rm I}$
to $\rho_{\rm II}$ (and insensitive to the details
of $T_{\rm cp}(\rho)$ outside this range).

(III) {\it Massive} NSs show 
{\it fast} cooling similar to the fast cooling of
non-superfluid NSs. These stars have central kernels where
the direct Urca process is either unaffected or weakly suppressed
by the proton superfluidity. In such kernels,
$T_{\rm cp}(\rho) \la T_{\rm cp}^{\rm (II)}$.
The central densities and masses of rapidly cooling NSs lie in the range
given by Eq.\ (\ref{MIII}).
The thermal evolution 
of rapidly cooling NSs is not very sensitive to
the model of $T_{\rm cp}(\rho)$ and to EOS
in the stellar core. Note that if $\rho_{\rm cmax}< \rho_{\rm II}$,
the rapidly cooling NSs do not exist.
In the frame of our interpretation, 
no NS observed so far can be assigned to this class.

\subsection{Crustal superfluidity and slow cooling}
\label{sect4-3}

As the next step, we retain proton superfluidity
and add $^1$S$_0$ neutron superfluidity 1ns in the NS crust
and outermost core.
In this case we obtain the solid curves 
in Figs.\ \ref{fig5}--\ref{fig7}. 
For the moderately or rapidly cooling middle-aged NSs 
($M > M_{\rm I}$) they are fairly close to 
the dot-and-dashed curves. This is quite expected
(e.g., Gnedin et al.\ \cite{gyp01}): the $^1$S$_0$
neutron superfluidity is mainly located in the NS crust which is much less
massive than the NS core. Thus,
the crustal superfluidity does not affect noticeably
our interpretation
of 1E 1207--43, RX J0002+62, Vela, PSR 0656+14, and Geminga
in terms of moderately cooling NSs.

However, as pointed out in Paper II, 
this crustal superfluidity strongly affects the slow cooling
of low-mass NSs ($M < M_{\rm I}$). 
Its effects are twofold. First, at $t \la 3 \times 10^5$ yr
the neutrino luminosity due to $^1$S$_0$ pairing of neutrons
may dominate    
the sufficiently low neutrino luminosity
of the stellar core (see Fig.\ \ref{fig4} and Paper II, for details). 
Second, at $t \ga 10^5$ yr the $^1$S$_0$ neutron superfluidity
reduces the heat capacity of the crust. Both effects
accelerate NS cooling and decrease $T_{\rm s}^\infty$ 
(Figs.\ \ref{fig5}--\ref{fig7})
violating our interpretation of the three sufficiently hot sources,
RX J0822--43, PSR 1055--52, and RX J1856--3754.
The interpretation of RX J1856--3754
is affected to a lesser extent, 
as a consequence of the rather large errorbar of $T_{\rm s}^\infty$
for this source  
(Sect.\ 3).

Let us demonstrate that our interpretation can be rescued by the
appropriate choice of $T_{\rm cns}(\rho)$.
For this purpose we focus on the interpretation of 
RX J0822--43, PSR 1055--52, and RX J1856--3754, as 
the very slowly cooling NSs ($M \leq M_{\rm I}$).

\begin{figure}
\centering
\epsfxsize=86mm
\epsffile[20 143 590 720]{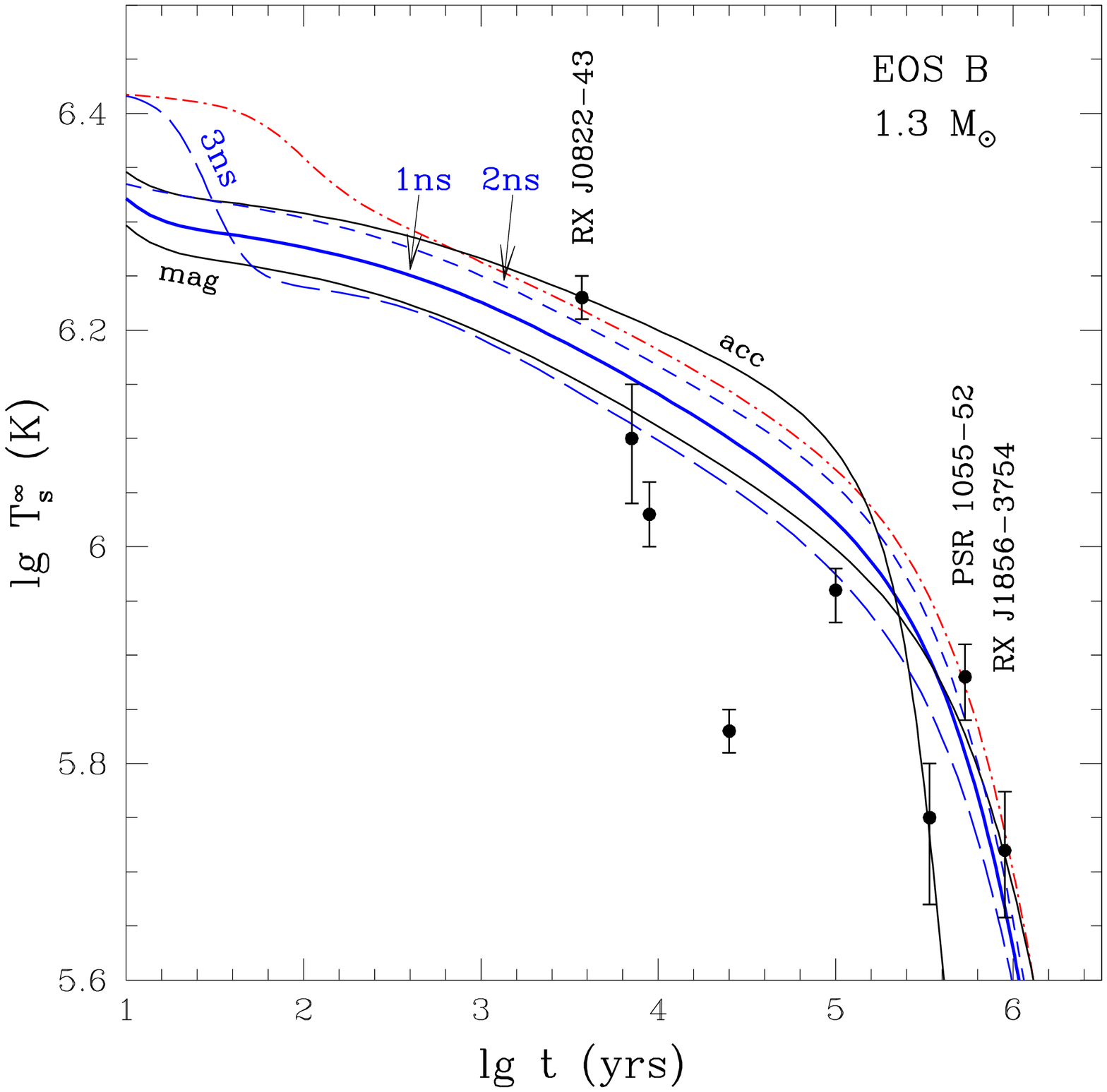}
\caption{
Cooling curves of $1.3 \, {\rm M}_\odot$ NS model
versus observations of RX J0822--43, PSR 1055--52,
and RX J1856--3754. Dot-and-dashed curve: 
proton superfluidity 3p in the NS core. Solid, short-dashed,
and long-dashed curves include, in addition,
models 1ns, 2ns, and 3ns of crustal neutron superfluidity, respectively.
Thick solid line is the same as
in Fig.\ \ref{fig7}. Thin solid curve {\it acc} is calculated
assuming the presence of $2 \times 10^{-11}\, {\rm M}_\odot$ of
hydrogen on the NS surface. Thin solid curve
{\it mag} is obtained assuming the dipole
surface magnetic field ($10^{12}$ G at the
magnetic pole).
}
\label{fig10}
\end{figure}

For certainty, let us take EOS B, $M=1.3 \, {\rm M}_\odot$,
and proton superfluid 3p. The results are presented in Fig.\ \ref{fig10}
(cf with Fig.\ 3 of Paper II). 
The dot-and-dashed line is the same as in Fig.\ \ref{fig7} and neglects
the crustal neutron superfluidity.
Thick solid line is also the same as in Fig.\ \ref{fig7}.
It includes an additional effect of
crustal superfluid 1ns and lies below the
observational limits on $T_{\rm s}^\infty$ for the sources in question
(or almost below in case of RX J1856--3754).
To keep the proposed interpretation of the three sources
we must raise the cooling curves calculated including the
crustal superfluidity. 
To this aim, we must suppress the neutrino emission associated with
$^1$S$_0$ pairing of neutrons (Fig.\ \ref{fig4}). 
Recall that in a middle-aged NS
this emission is mainly
generated (Fig.\ \ref{fig4}) in two relatively narrow layers,
near the neutron drip point and near the crust-boundary interface,
where the local NS temperature $T$ is just below $T_{\rm cns}(\rho)$.
Since the Cooper-pairing 
neutrino luminosity is roughly proportional
to the widths of these emitting  
layers,
we can achieve our goal by reducing their widths. 
This can be done by setting $T_{\rm cns}^{\rm max}$ 
higher and by making $T_{\rm cns}(\rho)$ decrease sharper
in the wings (see Paper II, for details). 

For example, taking crustal superfluid 2ns
instead of 1ns (Figs.\ \ref{fig1} and \ref{fig2}) we obtain the dashed
cooling curve in Fig.\ \ref{fig10} which comes much closer
to the dot-and-dashed curve than the thick solid curve (model 1ns).
(Another example: shifting $T_{\rm cns}(\rho)$ for model
2ns into the crust would additionally raise the curve
towards the dot-and-dashed one.) 
Note that the cooling curves
are insensitive to the details of $T_{\rm cns}(\rho)$
profile near the maximum,  
as long as
$T_{\rm cns}^{\rm max}\ga 5 \times 10^9$ K,
but they are extremely sensitive to the decreasing
slopes of $T_{\rm cns}(\rho)$. 
On the other hand, by taking the smoother and lower 
$T_{\rm cns}(\rho)$, model 3ns, we obtain
a colder NS than
needed for the interpretation of observations 
(long-dash line in Fig.\ \ref{fig10}).
Therefore, $^1$S$_0$ neutron
superfluidity with maximum $T_{\rm cns}^{\rm max} < 5 \times 10^9$ K
and/or with smoothly decreasing slopes of the $T_{\rm cns}(\rho)$
profile near the crust-core interface and
neutron drip point {\it violates}
the proposed interpretation of the observational data.    

Let us stress that the observations
of RX J0822--43, PSR 1055--52, and RX J1856--3754 
can be fitted even with our
initial model 1ns of the crustal neutron superfluidity. The
high surface temperature of RX J0822--43 can be explained
assuming additionally the presence of a low-mass
($2 \times 10^{-11} \, {\rm M}_\odot$) heat-blanketing 
surface envelope of hydrogen or helium.  
This effect is modeled using 
the results of Potekhin et al.\ (\cite{pcy97}).   
Light elements increase the electron thermal conductivity
of NS surface layers and raise $T_{\rm s}^\infty$ at the
neutrino cooling stage (curve {\it acc} in Fig.\ \ref{fig10}).
In order to explain the observations of PSR 1055--52
and RX J1856--3754, we can assume again 
model 1ns of crustal superfluidity,
iron surface and the dipole surface magnetic field
($\sim 10^{12}$ G at the magnetic pole;
line {\it mag} in Fig.\ \ref{fig10}).  
Such a field makes the NS surface layers
overall less heat-transparent (Potekhin \& Yakovlev \cite{py01}), 
rising $T_{\rm s}^\infty$
at $t \ga 3 \times 10^5$ yr. 
Note that the dipole field $\ga 10^{13}$ G
has the opposite effect, resembling the effect of
the surface envelope of light elements.
Therefore, we can additionally vary cooling
curves by assuming the presence of light elements
and/or the magnetic field on the NS surface. However,
these variations are less pronounced than those due to  
nucleon superfluidity. For instance, we cannot reconcile the cooling curves
with the observations of PSR 1055--52
assuming model 3ns of the crustal superfluidity     
with any surface magnetic field.

\subsection{$^3${\rm P}$_2$ pairing of neutrons in the NS core}
\label{sect4-4}

Now we focus on the effects of $^3$P$_2$ neutron
pairing, which were neglected so far. They are illustrated
in Fig.\ \ref{fig5}, as an example. They
would be qualitatively similar for the other cooling
models in Figs.\ \ref{fig6} and \ref{fig7}.
In Fig.\ \ref{fig5} we take the cooling models
obtained including proton superfluidity 1p
and crustal superfluidity 1ns and add the $^3$P$_2$
neutron superfluidity (model 1nt, Table 2) in the core.
We have the same
(solid) cooling curves for the young NSs
which have the internal temperatures $T$ above the maximum value of
$T_{\rm cnt}^{\rm max} \approx 3 \times 10^8$ K. However, when $T$
falls below $T_{\rm cnt}^{\rm max}$, we obtain 
(dots) a strong acceleration
of the cooling associated with the powerful neutrino emission
due to $^3$P$_2$ neutron pairing (Fig.\ \ref{fig4}).
This emission 
complicates our interpretation of
older sources, PSR 0656+14, Geminga, PSR 1055--52, and
RX J1856--3754. 
It may induce {\it really fast cooling} of such sources even if
their mass is {\it low}, $M < M_{\rm D}$ (Sect.\ 4.2).
To avoid this difficulty we assume
(Paper I) {\it weak} $^3$P$_2$
{\it pairing}, $T_{\rm cnt}(\rho)$, with maximum
$T_{\rm cnt}^{\rm max} < 10^8$ K. 
Then, it does not affect the proposed interpretation.

\section{Summary}
\label{sect-discuss}

Let us summarize the effects of the three types of superfluids on
NS cooling:

(a) Strong proton superfluidity in the NS cores,
combined with the direct Urca process at
$\rho > \rho_{\rm D}$, separates the cooling models into
three types: (I) slowly cooling, low-mass NSs
($M \la M_{\rm I}$); (II) moderately cooling, medium-mass NSs 
($M_{\rm I} \la M \la M_{\rm II}$);
(III) rapidly cooling, massive NSs ($M \ga M_{\rm II}$).
These models have distinctly
different properties (Sect.\ 4.2). The regime of moderate cooling
cannot be realized without the proton superfluidity.

(b) Strong proton superfluidity in the NS core is required to interpret
the observational data on the three sources, 
RX J0822--43, PSR 1055--52, and
RX J1856--3754, 
hot for their ages,   
as the very slowly cooling NSs
(Sects.\ 4.2, 4.3).
Within this interpretation,
all three sources may have masses
from about $1.1 \, {\rm M}_\odot$ to $M_{\rm I}$;
it would be difficult to determine their masses exactly 
or distinguish EOS in the NS core from the cooling models.

(c) Strong proton superfluidity is needed to interpret
observations of the other sources, 1E 1207--52,
RX J0002+62, Vela, PSR 0656+14, and Geminga,
as the medium-mass NSs.
This allows one to ``weigh'' these NSs, i.e., determine their
masses, for a given model of $T_{\rm cp}(\rho)$. 
The weighing is very sensitive to
the decreasing slope of $T_{\rm cp}(\rho)$ 
in the density range $\rho_{\rm I} \la  \rho  \la  \rho_{\rm II}$, and
it depends also 
on the EOS in the NS core (Sect.\ 4.2). 

(d) Strong or moderate $^3$P$_2$ neutron superfluidity
in the NS core initiates rapid cooling due to the neutrino emission
resulted from neutron pairing. This invalidates the proposed
interpretation of the old sources like PSR 0656+14,
Geminga, PSR 1055--52, and RX J1856--3754.
To save the interpretation, we assume a weak $^3$P$_2$
neutron superfluidity, $T_{\rm cnt}^{\rm max} < 10^8$ K (Sect.\ 4.4).

(e) $^1$S$_0$ neutron superfluidity in the crust
may initiate a strong Cooper-pairing neutrino emission, 
decrease substantially $T_{\rm s}^\infty$
of the slowly cooling NSs, and weaken
our interpretation
of RX J0822--43, PSR 1055--52, and RX J1856--3754 (although it does not
affect significantly the moderate or fast cooling).
We can save the interpretation by assuming
that the maximum of the critical temperature profile
$T_{\rm cns}(\rho)$ is not too small
($T_{\rm cns}^{\rm max} \ga 5 \times 10^9$ K) and the                     
profile decreases
sharply in the wings (Sect.\ 4.3).

(f) The interpretation of the slowly cooling sources
is sensitive to
the presence of the surface magnetic fields and/or
heat-blanketing surface layer composed
of light elements (Sect.\ 4.3). 

(g) No isolated middle-aged NSs 
observed so far can be identified as a
rapidly cooling NS. In the frame of our models, these NSs 
do not exist
for those EOSs and superfluid $T_{\rm cp}(\rho)$  
for which $M_{\rm max}< M_{\rm II}$.

If our interpretation is correct, we can make the following conclusions
on the properties of dense matter in NS interiors.   

(i) Strong proton superfluidity we need is in favor of
a not too large symmetry energy at supranuclear densities (Paper I).
A very large symmetry energy would mean a high proton fraction
which would suppress proton pairing. On the other hand,
the symmetry energy should not be too small to open the direct Urca
process at $\rho > \rho_{\rm D}$.

(ii) Weak $^3$P$_2$ 
neutron pairing is in favor of a not too soft EOS
in the NS core (Paper I). The softness would mean a strong attractive
nn interaction and, therefore, strong neutron pairing.

(iii) Specific features of the crustal neutron superfluidity
we adopt
are in favor of those microscopic theories which predict  
$T_{\rm cns}(\rho)$ profiles with $T_{\rm cns}^{\rm max}
\ga 5 \times 10^9$ K (or $\Delta_{\rm cn}(k)$ profiles
with $\Delta_{\rm cn}^{\rm max} \ga 1$ MeV, see Fig.\ \ref{fig2}).
This is in line with many microscopic calculations
of the superfluid gaps which include the medium polarization effects in
nn interaction (e.g., Lombardo \& Schulze \cite{ls01}). 
However, the reduction of the gap
by the medium polarization should not be too strong, and
the decreasing slope of $\Delta(k)$ 
should be rather sharp.
These requirements constrain the microscopic theories. 

The proposed interpretation of the observations
relates the inferred NS masses to the superfluid properties of NS interiors.
By varying EOS and the proton critical temperature,  
we can attribute different masses to the same sources.
If, on the other hand, we knew the range of masses
of the cooling middle-aged NSs we would be able to draw definite
conclusions on the superfluid state of their interiors,
first of all, on the proton critical temperature, $T_{\rm cp}(\rho)$.

Our analysis may seem too simplified because
we neglect a possible presence of other particles in the NS cores
(muons, hyperons, quarks). We expect that
the inclusion of other particles and the effects
of superfluidity of hyperons or quarks will
complicate theoretical analysis but will not
change our basic conclusion on the existence
of the slowly, moderately, and rapidly cooling NSs.  

Our calculations show that the cooling of middle-aged NSs
with $M < M_{\rm I}$ is sensitive to the density profile
of free neutrons near the crust bottom and
neutron drip point. We have used only one
model of the free-neutron distribution in the crust, assuming
the atomic nuclei to be spherical at the crust bottom.
It would be interesting to consider the models
of crust matter with non-spherical nuclei
(e.g., Pethick \& Ravenhall \cite{pr95})
and 
the effects of superfluidity of nucleons confined in the atomic nuclei
in the NS crust. 

Let us stress that determination of $T_{\rm s}^\infty$
from observational data is a very complicated problem 
(as described in part by Yakovlev et al.\ \cite{yls99}).
It requires very high-quality data and
theoretical models of NS atmospheres.
Thus, the current values of $T_{\rm s}^\infty$ may change
substantially after the forthcoming observations
and new theoretical modeling. These changes may affect 
our interpretation of the observational data,
first of all, of RX J0822--43, PSR 1055--52, and
RX J1856--3754. For instance, RX J1856--3754
may have a colder surface ($T_{\rm s}^\infty \sim 0.25$ MK),
than assumed in the above analysis,
with a hot spot (e.g., Pons et al.\ \cite{ponsetal01},
Burwitz et al.\ \cite{burwitzetal01},
G\"ansicke et al.\ \cite{gbr01}). If confirmed,
the lower $T_{\rm s}^\infty$ might be explained by
the effect of $^3$P$_2$ neutron pairing (Fig.\ \ref{fig5}).
We expect that future observations
of the thermal emission from these sources will be
crucial for understanding the superfluid properties
of NS matter. 

\begin{acknowledgements}
We are grateful to G.G.\ Pavlov for encouragement,
to M.E.\ Gusakov,
K.P.\ Levenfish, and A.Y.\ Potekhin for critical remarks, and to
P.\ Haensel, our coauthor of Paper I,
for useful comments.
One of the authors (DGY) is grateful to the Institute
for Nuclear Theory at the University of Washington
for its hospitality and to the Department of Energy for partial support
during the completion of this work.
The work was partially supported by RFBR (grant No.\ 99-02-18099).
\end{acknowledgements}

\end{document}